\documentclass[runningheads]{llncs}

\usepackage[T1]{fontenc}
\usepackage{xcolor}
\usepackage[colorlinks=true, linkcolor=blue, urlcolor=blue, citecolor=blue, bookmarks=true, bookmarksopen=true, bookmarksdepth=3]{hyperref}
\usepackage{bookmark}

\usepackage{tikz}
\usetikzlibrary{decorations.pathreplacing}

\usepackage{soul}
\usepackage{graphicx}
\usepackage{amsmath}
\usepackage{amssymb}
\usepackage{algorithm}
\usepackage[noend]{algpseudocode}
\usepackage{orcidlink}
\usepackage[normalem]{ulem}

\newcommand*{\QEDB}{\null\nobreak\hfill\ensuremath{\square}}
\newcommand{\bo}[1]{\textbf{\textit{{#1}}}}

\newcommand{\bigo}[0]{\mathcal{O}}

\def\s#1{\mbox{\boldmath $#1$}} 

\begin{document}
\title{On the Complexity of Finding Approximate LCS of
Multiple Strings}

\author{Hamed Hasibi\orcidlink{0000-0002-6371-2991} \and
Neerja Mhaskar\orcidlink{0000-0002-3233-6540} 
\and
W.\ F.\ Smyth\orcidlink{0000-0002-6455-7751}
}

\authorrunning{H. Hasibi \textit{et al.}}

\institute{Department of Computing \& Software, McMaster University, Canada
\\
\url{{hasibih, pophlin, smyth}@mcmaster.ca} 
}

\maketitle              
\begin{abstract}

Finding an Approximate Longest Common Substring (ALCS) within a given set $\s{S}=\{s_1,s_2,\ldots,s_m\}$ of $m \ge 2$ strings is a key problem in computational biology, such as identifying related mutations across multiple genetic sequences.
We study several variants of ALCS problems that, given integers $k$ and $t \le m$, seek the longest string $u$ --- or the longest substring $u$ of any string in $\s{S}$ --- that lies within distance $k$ of at least one substring in $t$ distinct strings from $\s{S}$. While the general problems are NP-hard, we present efficient algorithms for restricted cases under Hamming and edit distances using the $LCP_k$ and $k$-errata tree data structures. Our methods achieve run times of $\mathcal{O}(N^2)$, $\mathcal{O}(k\ell N^2)$, and $\mathcal{O}(mN\log^k \ell)$, where $\ell$ is the length of the longest string and $N$ is the sum of the lengths of all the strings in $\s{S}$. We also establish conditional lower bounds under the Strong Exponential Time Hypothesis and extend our study to indeterminate strings.

\keywords{Approximate Longest Common Substring \and Hamming Distance \and Edit Distance \and k-errata Trie \and Strong Exponential Time Hypothesis \and Indeterminate String \and Degenerate string \and Matrix Multiplication}
\end{abstract}

\section{Introduction}\label{sec:intro}

Finding similar substrings occurring in strings from a given string set is a challenging problem in many application areas, particularly in the analysis of genomic sequences in computational biology. Identifying the Longest Common Substring (LCS) of two strings or a collection of strings is widely used to quantify similarity among DNA sequences (see~\cite[Chapter~7]{gusfield1997algorithms},~\cite{ulitsky2006average}). Due to evolutionary processes affecting biological sequences, exact common substrings may not always be present. As a result, ``approximate'' occurrences --- such as the Approximate Longest Common Substring (ALCS) --- are of significant interest. Moreover, exact applications of the LCS problem are not always useful for measuring the similarity between two or more sequences.
To more accurately address these similarities, several problems related to the ALCS of two strings have been defined~\cite{Abboud2015MoreDesign,Babenko2011ComputingMismatch,Flouri2015LongestMismatches,Grabowski2015AProblem,Thankachan2016AProblem}. In this paper, we study ALCS of more than two strings.

In Section~\ref{sec:prelimin}, we introduce the necessary terminology. Section~\ref{sec:DefforApproxLCS} defines new problems related to the ALCS of multiple strings. In Section~\ref{sec:litReview}, we review existing literature on LCS.
In Section~\ref{sec:main}, we show that the general ALCS problems are NP-hard, and present polynomial time and polylog time solutions to their restricted variants, while Section~\ref{sec:indet} explores these problems in the context of indeterminate strings.
Finally, Section~\ref{sec:seth1} presents SETH-hardness results.

\section{Preliminaries}\label{sec:prelimin}
A \bo{string} $s$ is a sequence of $n \geq 0$ \bo{letters} drawn from a finite ordered set $\Sigma$ called an \bo{alphabet}. Alphabet \bo{size} is $\sigma = |\Sigma|$. Thus, we treat a string $s$ of length $n$ as an array $s[1..n]$ of elements drawn from $\Sigma$. We define $\Sigma^*$ as the set of all strings over the alphabet $\Sigma$, and define $\Sigma^+$ as the set of all non-empty strings over $\Sigma$. An \bo{empty} string $(n=0)$ is denoted by $\s{\varepsilon}$.
For any pair of integers $i$ and $j$ satisfying $1 \leq i \leq j \leq n$, we define a \bo{substring} $s[i..j]$ (sometimes called a \bo{factor}) of $s$ as follows: $s[i..j]=s[i]s[i+1]..s[j]$. We say that substring $s_1[i..j]$ of $s_1$ \bo{occurs} in string $s_2$ if  
$s_2[i'..j'] = s_1[i..j]$, where $1 \leq i' \leq j' \leq |s_2|$. A \bo{prefix} (\bo{suffix}) of a nonempty $s$ is a substring $s[i..j]$, where $i=1$ $(j=n)$.

An \bo{edit operation} on a string is the insertion, deletion or substitution of a single letter.
The \bo{cost} of an edit operation is an integer function of the operation type and the letters involved.
Given two strings $s_1$ and $s_2$, both of length $n$, the \bo{Hamming distance} $d_H(s_1,s_2)$, is defined to be the minimum cost of substitutions that transform $s_1$ to $s_2$. For any two strings $s_1$ and $s_2$ of arbitrary lengths, the \bo{edit distance} (or \bo{Levenshtein distance})
$d_{E}(s_1, s_2)$ is
the minimum cost over all sequences of unit cost edit operations that transform $s_1$ to $s_2$.
When the edit operations are assigned different (non-unit) costs, the resulting measure is called the \bo{weighted edit distance}, denoted by $d_{w}(s_1, s_2)$.
A \bo{k-approximate occurrence} of any string $u$ under distance $\delta \in \{H,E,W\}$ is a string $u'$ such that $d_{\delta}(u,u') \leq k$.
An \bo{indeterminate letter} is a non-empty subset of $\Sigma$ of cardinality greater than one. Let $\Sigma'$  be the set of all non-empty subsets of $\Sigma$. An \bo{indeterminate} (or \bo{degenerate}) string $\Tilde{s}[1..|s|]$ is an array of letters drawn from $\Sigma'$ such that at least one letter is indeterminate. In $\Tilde{s}$, letters $\Tilde{s}[i]$ and $\Tilde{s}[j]$ are said to \bo{match} if and only if $\Tilde{s}[i] \cap \Tilde{s}[j] \neq \emptyset$, $1 \leq i,j \leq |\Tilde{s}|$, and we write $\Tilde{s}[i] \approx \Tilde{s}[j]$. Otherwise, they \bo{mismatch}, and we write $\Tilde{s}[i] \not\approx \Tilde{s}[j]$.

\section{Definitions of ALCS Problems on Multiple Strings}\label{sec:DefforApproxLCS}

We now define problems over $\Sigma$ to compute the Approximate Longest Common Substring (ALCS) across multiple strings, where the distance allowed is bounded by a constant $k \in \mathbb{N}$ with $k>0$. Specifically,
for given finite values $k, t, m \in \mathbb{N}$, $t\leq m$, and a set $\s{S}=\{s_1, s_2, \ldots , s_m\}$ of non-empty strings, where $\ell$ is the length of the longest string in $\s{S}$, we define the following problems:
\begin{enumerate}
    \item \textbf{\textit{kt} Longest Common Substring (\textit{kt}-LCS)}: find a longest string $u \in \Sigma^+$ such that there exists a subset $\s{S}' = \{s'_1,s'_2,\ldots,s'_t\} \subseteq \s{S}$, for which substrings $u'_i$ of $s'_i$, $1 \le i \leq t$, satisfy $d_\delta(u,u'_i) \le k$.

    \item \textbf{Restricted \textit{kt} Longest Common Substring (\textit{Rkt}-LCS)}: find a longest non-empty substring $u$ of any string in \s{S} such that there exists a subset $\s{S}' = \{s'_1,s'_2,\ldots,s'_t\} \subseteq \s{S}$, for which substrings $u'_i$ of $s'_i$, $1 \le i \leq t$, satisfy $d_\delta(u,u'_i) \le k$.

    \item \textbf{Restricted \textit{k}-LCS Set (\textit{Rk}-LCSS):} find a set of longest substrings $\s{U}=\{u_1,\ldots,u_m\}$ of $\s{S}=\{s_1,\ldots,s_m\}$, such that $d_\delta(u_i,u_j) \leq k$ for any $1\leq i,j \leq m$, and each $u_i$ is a  substrings of $s_i \in \s{S}$.
\end{enumerate}

When $t = m$, we refer to the $kt$-LCS and $Rkt$-LCS problems as \textbf{\textit{k}-LCS} and \textbf{\textit{Rk}-LCS} problems, respectively. For $k=0$, the $Rkt$-LCS and $kt$-LCS  problems are equivalent. For $\s{S}=\{s_1,s_2,s_3,s_4\} = \{aabcf,fabcd,dgiabc,ahabch\}$ with $k=2$ and $t=3$, we find substring $iabc$ of $s_3$ is the $Rk$-LCS of \s{S}, while substring $aabcf$ of $s_1$ is the $Rkt$-LCS of \s{S}.

\section{Literature Review}\label{sec:litReview}

In this section, we review the three following categories of LCS problems. These problems can be regarded as variations of the problems defined in Section~\ref{sec:DefforApproxLCS}: 

\begin{itemize}
    \item ELCS of two strings \textit{(\textit{Rk}-LCS, where \textit{k=0} and \textit{m=2})} 
    \item ELCS of more than two strings 
    \textit{(\textit{Rkt}-LCS, where \textit{k=0} and $2 < t \leq m$)}
    \item ALCS of two strings \textit{(\textit{Rk}-LCS, where $k> 0$ and \textit{m=2})}
\end{itemize}

\textbf{ELCS of two strings.} Given two strings $s_1$ and $s_2$, each of length $n$, find a longest substring occurring in both $s_1$ and $s_2$.
It has been shown that this problem can be solved in $\bigo(n)$-time ~\cite{Weiner1973LinearAlgorithms,Farach1997OptimalAlphabets}. Using the results in~\cite{Farach1997OptimalAlphabets}, Charalampopoulos \textit{et al.} later showed that the ELCS problem over an alphabet $[0,\sigma)$ can be solved in $\bigo(n \log \sigma/ \sqrt{\log n})$-time and $\bigo(n/ \log_\sigma n)$-space complexity, where $\log \sigma = o(\sqrt{\log n})$~\cite{Charalampopoulos2021FasterSubstring}.
\textbf{ELCS of more than two strings.} Given a string set $\s{S}=\{s_1, s_2, \ldots , s_m\}$, $m$ positive integers $x_1,x_2,..,x_m$, and $t$ where $1 \leq t \leq m$, find a longest substring $u$ of any string in \s{S} for which there are at least $t$ strings $s_{i_1},s_{i_2},\ldots,s_{i_t}$ $(1 \leq i_1 < i_2 <\ldots< i_t \leq m)$ such that $u$ occurs at least $x_{i_j}$ times in $s_{i_j}$ for each $j$ with $1 \leq j \leq t$~\cite{Arnold2011LinearProblem}. Chi and Hui~\cite{ChiColor}, and Lee and Pinzon~\cite{Lee2007ARepeats} show that this problem can be solved in $\bigo(N)$ time, where $N=\sum_{i=1}^{m}|s_{i}|$. A simple solution for all $1 \leq t \leq m$, requires $\bigo(mN)$ time using Lee and Pinzon results. However, a much simpler and more efficient solution was proposed by Arnold and Ohlebusch that requires $\bigo(N)$ time for all $1 \leq t \leq m$~\cite{Arnold2011LinearProblem}.

\textbf{ALCS of two strings for $\delta=H$}. Given two strings $s_1$ and $s_2$, each of length $n$, find a longest substring (of either $s_1$ or $s_2$) that occurs in both $s_1$ and $s_2$ with at most $k$ mismatches. 
In 2011, Babenko and Starikovskaya~\cite{Babenko2011ComputingMismatch} solve this problem for $k=1$ in $\bigo (n^2)$-time and $\bigo (n)$-space complexity. This result for $k=1$ was later improved by Flouri \textit{et al.} in~\cite{Flouri2015LongestMismatches}, where they propose an $\bigo(n \log n)$-time and $\bigo(n)$-space approach. 
Leimeister and Morgenstern, in 2014, introduced this problem for $k>1$ and proposed a greedy heuristic algorithm to solve it~\cite{Leimeister2014Kmacs:Comparison}. Later in 2015, Flouri \textit{et al.}~\cite{Flouri2015LongestMismatches} presented a simple and elegant $\bigo(n^2)$-time algorithm that uses constant additional space. 

Grabowski~\cite{Grabowski2015AProblem} presented two output-dependent algorithms, with $\bigo (n((k+1)(l_0+1))^k)$ and $\bigo(n^2k/l_k)$ time complexities, where $l_0$ is the length of the ELCS of $s_1$ and $s_2$ and $l_k$ is the length of the ALCS of $s_1$ and $s_2$. Abboud \textit{et al.} in~\cite{Abboud2015MoreDesign} described a $k^{1.5}n^2/2^{\Omega(\sqrt{(\log n)/k})}$-time randomized solution to this problem. In 2016, Thankachan \textit{et al.}~\cite{Thankachan2016AProblem} presented an $\bigo (n \log^k n)$-time and $\bigo (n)$-space solution. Then in 2018, Charalampopoulos \textit{et al.} proved an $\bigo (n)$-time and space solution where the length of $ALCS$ is $\Omega(\log^{2k+2} n)$~\cite{Charalampopoulos2018Linear-timeMismatches}. In 2019, Kociumaka \textit{et al.} showed that assuming the Strong Exponential Time Hypothesis (SETH)~\cite{Impagliazzo2001OnK-SAT}\cite{Impagliazzo2001WhichComplexity}, no strongly subquadratic-time solution for this problem exists for $k \in \Omega(\log n)$~\cite{Kociumaka2019LongestMismatches}. 
Finally in 2021, Charalampopoulos \textit{et al.}~\cite{Charalampopoulos2021FasterSubstring} showed that this problem can be solved in $\bigo (n \log^{k-1/2} n)$-time and $\bigo (n)$-space complexity. 

\textbf{ALCS of two strings for $\delta \in \{E,W\}$}. Given two strings $s_1$ and $s_2$, each of length $n$, find a longest substring (of either $s_1$ or $s_2$) that occurs in both $s_1$ and $s_2$ with at most $k$ edit operations.
In 2015 Abboud \textit{et al.} described a $k^{1.5}n^2/2^{\Omega(\sqrt{(\log n)/k})}$-time randomized solution~\cite{Abboud2015MoreDesign} for ALCS of two strings for $\delta \in \{E,W\}$. In 2018, Thankachan \textit{et al.} proposed an $\bigo (n \log^k n)$-time and $\bigo (n)$-space solution to this problem~\cite{thankachan2018algorithmic}.

\section{Results on ALCS Problems on Multiple Strings}\label{sec:main}

The $k$-LCS problem for $\delta=H$ (defined in Section~\ref{sec:DefforApproxLCS}) is called the \textit{Longest Common Approximate Substring} problem in~\cite{smith2004common}, and has been proved to be NP-hard by reducing the \textit{Vertex Cover} problem to it. It has also been shown that this problem cannot be approximated in polynomial time with a performance ratio better than $2-\varepsilon$, for any $\varepsilon > 0$, unless $P=NP$ \cite[Theorems 4.12 \& 4.13]{smith2004common}. Since $k$-LCS is a special case of $kt$-LCS, we have:

\begin{theorem}
    The $kt$-LCS problem is NP-hard for $\delta=H$.
\end{theorem}

\subsection{Polynomial Time Solution for \texorpdfstring{$Rk$-LCS}{Rk-LCS} \& \texorpdfstring{$Rkt$-LCS}{Rkt-LCS} 
}\label{sec:poly}
In this section, we provide polynomial-time solutions to $Rk$-LCS and $Rkt$-LCS problems under different distance metrics. For this, we use an auxiliary table called \textit{lengthStat} (Definition~\ref{def:ls}). For strings $s_i$ and $s_j$, and $k \in \mathbb{N}$, $LCP^{H,k}_{(s_i,s_j)}[i',j']$,
$1\leq i' \leq |s_i|$ and $1\leq j' \leq |s_j|$, is the length of the longest common prefix, with at most $k$ mismatches, of suffixes $s_i[i'..|s_i|]$ and $s_j[j'..|s_j|]$. An example of $LCP^{H,k}_{(s_i,s_j)}$ is given in Table~\ref{fig:lcpkexample} of the Appendix.

\begin{definition}[LengthStat Data Structure]\label{def:ls} For $S=\{s_1,s_2,\ldots,s_m\}$, $1 \leq i,j \leq m$, $1 \leq p \leq |s_i|$, and $1 \leq l \leq |s_i| -p +1$, $lengthStat^{\delta,k}_{(p,i)}[l,j] = 1$ iff $s_i[p..p+l-1]$ has at least one $k$-approximate occurrence in $s_j$ under distance metric $\delta$; otherwise, $lengthStat^{\delta,k}_{(p,i)}[l,j] = 0$. The last $(m+1)$-th column contains the sum of the values in each row.
\end{definition}

Below is the equation under the Hamming distance; that is, $\delta = H$ metric. An example of $lengthStat^{H,k}_{(p,i)}$ is given in Table~\ref{tab:lengthstat} of the Appendix.

\begin{multline}
\label{eq:lengthStatLCP}
\mathsf{lengthStat}^{H,k}_{(p,i)}[l,j] =
\begin{cases}
  1, & \text{if }
       \begin{aligned}[t]
         \exists\, q \in [1, |s_j|] \text{ such that } 
         \mathsf{LCP}^{H,k}_{(s_i,s_j)}[p,q] \geq l
       \end{aligned} \\
  0, & \text{otherwise}
\end{cases} \\
\text{where } 1 \leq p \leq |s_i|,\quad 1 \leq l \leq |s_i| - p + 1,\quad 1 \leq i,j \leq m
\end{multline}

We propose Algorithm~\ref{alg:lengthStat} to compute the $lengthStat$ table for the suffix $s_i[p..|s_i|]$, and $\delta = H$. To align the result with the definition, line 7 performs a bitwise \textit{OR} operation across the rows. 
The following Lemma follows directly from the results in~\cite{Flouri2015LongestMismatches}:

\begin{algorithm}
\caption{Computes $lengthStat^{H,k}$ table for $s_i[p..|s_i|]$}
\label{alg:lengthStat}
\begin{algorithmic}[1]
\Procedure{lengthStat}{$LCP^{H,k}_{(s_i,s_1)}..LCP^{H,k}_{(s_i,s_m)},p,i$}
\State $lengthStat^{H,k}_{(p,i)}[1..|s_i|-p+1,1..m+1] \leftarrow 0$    \Comment{Initial table with all zeros}
\For{all $j \in [1..m]$ and $q \in [1..\ell]$} 

    \State $l \leftarrow LCP^{H,k}_{(s_i,s_j)}[p,q]$
    \State $lengthStat^{H,k}_{(p,i)}[l,j] \leftarrow 1$
    
\EndFor 
\For{$r$ from $|s_i|-p+1$ to $2$} 
    \State $lengthStat^{H,k}_{(p,i)}[r-1,1..m] \leftarrow lengthStat^{H,k}_{(p,i)}[r,1..m] \lor lengthStat^{H,k}_{(p,i)}[r-1,1..m]$
    \State $lengthStat^{H,k}_{(p,i)}[r,m+1]= \Sigma^m_{j=1} lengthStat^{H,k}_{(p,i)}[r,j]$

\EndFor
\State $lengthStat^{H,k}_{(p,i)}[1,m+1]= \Sigma^m_{j=1} lengthStat^{H,k}_{(p,i)}[1,j]$
 
\State \textbf{return} $lengthStat^{H,k}_{(p,i)}$
\EndProcedure
\end{algorithmic}
\end{algorithm}

\begin{lemma}\label{lem:lcp}
For a given set $\s{S}=\{s_1,\ldots,s_m\}$ of strings and a specific string $s_i \in \s{S}$, $LCP^{H,k}_{(s_i,s_j)}$ for all values of $j$ such that $1 \leq j \leq m$, can be computed in $\bigo(m\ell^2)$ time and $\bigo(m\ell^2)$ space.
\end{lemma}

\begin{lemma}\label{lem:lengthstat}
   
   Given $LCP^{H,k}_{(s_i,s_j)}$ tables for a fixed string $s_i \in \s{S}$ and for all values of $j$, such that $1 \leq j \leq m$, $lengthStat^{H,k}_{(p,i)}$ for the suffix $s_i[p..|s_i|]$ can be computed in $\bigo(N)$ time and space.

\end{lemma}
\begin{proof}
The \texttt{for} loop in line 3 of Algorithm~\ref{alg:lengthStat} takes $\bigo(m \ell)$ or $\bigo(N)$ time to iterate through all the $LCP^{H,k}_{(s_i,s_j)}[p,q]$ values, where $1 \leq j \leq m$ and $1 \leq q \leq \ell$. The \texttt{for} loop in line 6 takes $\bigo(\ell)$ time. Since reading $LCP^{H,k}$ values takes constant time, the total time and space complexity for computing $lengthStat^{H,k}_{(p,i)}$ for the suffix $s_i[p..|s_i|]$ is $\bigo(N)$. \QEDB
\end{proof}

We compute the \textit{Rk}-LCS or \textit{Rkt}-LCS using a greedy approach. A longest substring of string $s_i$ that satisfies the \textit{Rk}-LCS or \textit{Rkt}-LCS criteria is called a \textit{candidate} solution $C_i$. Thus, we obtain a set of $m$ candidate solutions $\{C_1,C_2,\ldots,C_m\}$. To compute a candidate solution $C_i$, we construct the $lengthStat^{H,k}_{(p,i)}$ tables for every $1 \leq p \leq \ell$. By Lemma~\ref{lem:lcp} and~\ref{lem:lengthstat}, this requires $\bigo(m \ell^2)$ time and $\bigo(m\ell^2)$ space, including the time and space needed to compute all the $LCP^{H,k}$ tables. Using the last column of $lengthStat_{(p,i)}$ tables, we pick the substring $C_i=s_i[p..p+l'-1]$ with maximum length $l'$ that has $k$-approximate occurrences in $m$ ($t$) distinct strings for \textit{Rk}-LCS (\textit{Rkt}-LCS). After computing each $C_i$, we can release the $\bigo(m \ell^2)$ space that is used by $lengthStat^{H,k}_{(p,i)}$ tables for every $1 \leq p \leq \ell$ and reuse it for the next candidate $C_{i+1}$. Computing all candidates requires $\bigo(m.m\ell^2)$ or $\bigo(N^2)$ time. Finally, we return a \textit{longest} candidate as our solution to the $Rk$-LCS (or $Rkt$-LCS). Thus, we get the following result.

\begin{theorem}\label{theo:complexity}
    The Rk-LCS and Rkt-LCS problems for $\s{S}=\{s_1, s_2, \ldots , s_m\}$ and $\delta=H$ can be computed in $\bigo(N^2)$ time and $\bigo(m \ell^2)$ additional space.
\end{theorem}

Now we provide solutions to the $Rk$-LCS and $Rkt$-LCS problems for $\delta \in \{E,W\}$. For this, we use the \textit{Longest Approximate Prefix} data structure introduced by K{\c{e}}dzierski and Radoszewski in~\cite{Kedzierski2022K-ApproximateDistance}:

\begin{definition}[Longest Approximate Prefix]\label{def:lap} Given strings $s_1$ and $s_2$, distance metric $\delta \in \{E,W\}$, and $k \in \mathbb{N}$, compute $P^{\delta,k}_{(s_1,s_2)}$ such that $P^{\delta,k}_{(s_1,s_2)}[a,b,a']$ is the maximum $b'\geq a'-1$, where $d_{\delta}(s_1[a,b],s_2[a',b']) \leq k$, or $-1$ if no such $b'$ exists.

\end{definition}
\begin{lemma}\label{lem:ptable}
    (\cite{Kedzierski2022K-ApproximateDistance}) For two strings $s_1$ and $s_2$, each of length at most $\ell$, $P^{\delta,k}_{(s_i,s_j)}$ table, where $\delta \in \{E,W\}$ is computed in $\bigo(k\ell^3)$  and $\bigo(\ell^3)$ space.
\end{lemma}
\begin{lemma}
    Computing the $P^{\delta,k}_{(s_i,s_j)}$ table for $\delta \in \{E,W\}$ and fixed $i$, for all $j$ values, $1 \leq j \leq m$, takes $\bigo(km\ell^3)$ time and $\bigo(m\ell^3)$ space.
\end{lemma}

For $\delta \in \{E,W\}$, we compute $lengthStat^{\delta,k}_{(p,i)}$ using $P^{\delta,k}_{(s_i,s_j)}$ tables for each suffix of any string in set \s{S}, similar to the computation shown for $\delta=H$.

\begin{multline}
\label{eq:lengthStatlev}
\mathsf{lengthStat}^{\delta,k}_{(p,i)}[l,j] =
\begin{cases}
  1, & \text{if }
       \begin{aligned}[t]
         \exists\, q \in [1, |s_j|] \text{ such that } 
         P^{\delta,k}_{(s_i,s_j)}[p, p+l-1, q] \neq -1
       \end{aligned} \\
  0, & \text{otherwise}
\end{cases} \\
\text{where } 1 \leq p \leq |s_i|,\quad 1 \leq l \leq |s_i| - p + 1,\quad 1 \leq i,j \leq m,\quad \delta \in \{E,W\}
\end{multline}

\begin{lemma}
        Given $P^{\delta,k}_{(s_i,s_j)}$ tables for a fixed string $s_i \in \s{S}$ and for all values of $j$, such that $1 \leq j \leq m$, $lengthStat^{\delta,k}_{(p,i)}$ for $\delta \in \{E,W\}$ for the suffix $s_i[p..|s_i|]$ can be computed in $\bigo(\ell N)$ time and $\bigo(N)$ space.
        
\end{lemma}
\begin{proof}
    For a fixed $i$ and $p$, we read values of $p^{\delta,k}_{(s_i,s_j)}[p,p+l-1,q]$ for $p+l-1,q \in \bigo(\ell)$ and $1 \leq j \leq m$. This requires $\bigo(m.\ell.\ell)$ or $\bigo(\ell N)$ time. Space requirement remains $\bigo(N)$. \QEDB
\end{proof}

\begin{theorem}\label{theo:le}
    The $Rk$-LCS and $Rkt$-LCS problems for $\delta \in \{E,W\}$ and $\s{S}=\{s_1, s_2, \ldots , s_m\}$ can be computed in $\bigo(k \ell N^2)$ time  and $\bigo(m\ell^3)$ space.
\end{theorem}
\begin{proof}

Similar to the case when $\delta = H$, computing each candidate $C_i$, for $1 \leq i \leq m$, requires calculating $lengthStat^{\delta,k}_{(p,i)}$ for every $1 \leq p \leq \ell$. For each candidate, the computation takes $\mathcal{O}(\ell \cdot \ell N + km\ell^3)$ time for generating the $lengthStat^{\delta,k}$ and $P^{\delta,k}$ tables, respectively, and requires $\mathcal{O}(m\ell^3)$ space.
For all $m$ candidates, the total time complexity is 
$\mathcal{O}(\ell^3m^2 + km^2\ell^3) $ or $ \mathcal{O}(k\ell N^2)$. In terms of space, we reuse the $\mathcal{O}(m\ell^3)$ space for each candidate. \QEDB
\end{proof}

\subsection{Polylog Time Solution for $Rk$-LCS}\label{sec:polylog}

Thankachan \textit{et al.} describe a $\bigo(\ell \log^k \ell)$-time and $\bigo(\ell)$-space solution for ALCS of two strings, at most of length $\ell$, for $\delta=H$~\cite{Thankachan2016AProblem} and $\delta=E$~\cite{thankachan2018algorithmic}. Their method uses \textit{k}-errata trie data structure proposed by Cole \textit{et al.} in~\cite{Cole2004DictionaryCares}.
Let $LCP^{E,k}_{(s_i,s_j)}[i',j']$ be the length of the longest common prefix between the suffixes $s_i[i'..|s_i|]$ and $s_j[j'..|s_j|]$, with at most $k$ edit operations. $MaxLCP^{H,k}_{(s_i,s_j)}[i']$ ($MaxLCP^{E,k}_{(s_i,s_j)}[i']$) is an array of size $|s_i|$, where each element $i'$ contains the maximum value of $LCP^{H,k}_{(s_i,s_j)}[i',j']$ ($LCP^{E,k}_{(s_i,s_j)}[i',j']$) for all $1 \leq j' \leq |s_j|$.

\begin{theorem}\label{theo:lcpk*}
    (\cite{Thankachan2016AProblem},\cite{thankachan2018algorithmic}) Let $s_i$ and $s_j$ be strings of length at most $\ell$. $MaxLCP^{H,k}_{(s_i,s_j)}[i']$ and $MaxLCP^{E,k}_{(s_i,s_j)}[i']$ arrays can be computed in $\bigo(\ell \log^k \ell)$ overall time using $\bigo(\ell)$ space for \textbf{all} values of $i'$.
\end{theorem}

\begin{theorem}
    For any $\delta \in \{H,E\}$, the $Rk$-LCS problem can be computed by solving $T=\max\limits_{1 \leq i \leq m} \max\limits_{1 \leq i' \leq |s_i|} \min\limits_{1 \leq j \leq m} MaxLCP^{\delta,k}_{(s_i,s_j)}[i']$.
\end{theorem}
\begin{proof}
For a string $s_i$ and its suffix $s_i[i'..|s_i|]$, $\min\limits_{1 \leq j \leq m} MaxLCP^{\delta,k}_{(s_i,s_j)}[i']$ is the length of the longest prefix of $s_i[i'..|s_i|]$ that occurs in all $\s{S}-\{s_i\}$ strings with at most $k$ errors. $T=\max\limits_{1 \leq i \leq m} \max\limits_{1 \leq i' \leq |s_i|} \min\limits_{1 \leq j \leq m} MaxLCP^{\delta,k}_{(s_i,s_j)}[i']$ is the length of a $Rk$-LCS of set \s{S}. We need to find a pair $(i,i')$ that maximizes $T$. Argument pair $(i,i')$ maximizing $T$ corresponds to $i'$-the suffix of $s_i$ such that substring $s_i[i'..i'+T-1]$ is a $Rk$-LCS.
We prove the correctness of this solution by contradiction. Assume that pair $(i,i')$ is not a $Rk$-LCS and there exists a pair $(v,v')$ which is an answer to this problem. Therefore, the $T$ value for $(v,v')$ is larger than its value for $(i,i')$; however, pair $(v,v')$ has not been captured by $T$. This is in contradiction with maximizing $T$. \QEDB
\end{proof}

\begin{theorem}\label{theo:klcslcpk*}
    The Rk-LCS problem for $\s{S}=\{s_1, s_2, \ldots , s_m\}$ can be computed in $\bigo(mN\log^k \ell)$ time with $\bigo(N)$ additional space for any $\delta \in \{H,E\}$. 
\end{theorem}

\begin{proof}
    The time complexity for computing $MaxLCP^{\delta,k}_{(s_i,s_j)}[i']$ for all $i'$, $i$, and $j$ values (string pairs) is $\bigo(m^2 \ell \log^k \ell) $ or $ \bigo(mN \log^k \ell)$. This is achieved by calculating $MaxLCP^{\delta,k}_{(s_i,s_j)}[i']$ for $m^2$ string pairs. Finding argument pair $(i,i')$ of $T$ takes $\bigo(m^2 \ell) $ or $\bigo(mN)$ time, which is less than the time complexity of computing $MaxLCP^{\delta,k}_{(s_i,s_j)}[i']$ for all $i'$, $i$, and $j$ values, Thus, the overall time complexity of the $Rk$-LCS problem under a distance metric in $\{H,E\}$ is $\bigo(mN \log^k \ell)$. 
    For a given string $s_i$, $MaxLCP^{\delta,k}_{(s_i, s_j)}[i']$ must be computed for all $1 \leq j \leq m$, $j \neq i$. This requires  $\bigo(m\ell) $ or $ \bigo(N)$ space. Since this space can be reused for other values of $i$, the overall space complexity remains $\bigo(N)$.  \QEDB  
\end{proof}

The $\bigo(mN\log^k \ell)$-time solution of the $Rk$-LCS problem outperforms the $\bigo(N^2)$-time solution for $\delta=H$ when $k< \frac{\log \ell}{\log \log \ell}$, and the $\bigo(k \ell N^2)$-time solution for $\delta=E$ when $k< \frac{\log k+ 2\log \ell}{\log \log \ell}$. By examining all $\binom{m}{t}$ set of strings, the $Rkt$-LCS problem can be computed in $\bigo(\binom{m}{t} mN\log^k \ell)$ time.

\subsection{$Rkt$-LCS of Multiple Indeterminate Strings for $\delta=H$}\label{sec:indet}
Now we explore the $Rk$-LCS and $Rkt$-LCS problems for $\delta=H$ for indeterminate strings. Analogous to $LCP^{H,k}_{(s_1,s_2)}$, we define $LCP^{H,k}_{(\Tilde{s}_1,\Tilde{s}_2)}$ and propose its efficient computation. For this, we adapt the approach presented by Flouri et al. to compute $LCP^{H,k}_{(\Tilde{s}_1,\Tilde{s}_2)}$, by simply replacing $\neq$ with $\not\approx$ in line 10 of the pseudocode given in~\cite[Figure 1]{Flouri2015LongestMismatches}. Assuming that the characters in indeterminate letter are sorted,  the intersection operation can be achieved in $\bigo(\sigma)$ time. Hence, a naive approach to computing $LCP^{H,k}_{(\Tilde{s}_i \Tilde{s}_j)}$ requires $\bigo(\sigma \ell^2)$-time. 
Using the approach presented in~\cite[Section 3.1]{iliopoulos2009algorithms} and fast matrix multiplication, we can determine the matches between every pair of indexes in $\Tilde{s}_i$ and $\Tilde{s}_j$ using the $\mathcal{I}[i',j']$ table, $1 \leq i'\leq |\Tilde{s}_i|$, $1 \leq j'\leq |\Tilde{s}_j$|, in $\bigo(\sigma\ell+\sigma^{w-2}\ell^2)$ time and $\ell^2$ space, where $w$ is the matrix multiplication
exponent\footnote{The best current known $w$ is $2.371866$~\cite{duan2023faster}.}. Hence, after computing the matrix $\mathcal{I}[i',j']$, checking matches between any pair of indeterminate letters of the strings $\Tilde{s}_i$ and $\Tilde{s}_j$ can be done in constant time.

\begin{equation}
\mathcal{I}[i',j'] =
\begin{cases}
1 & \text{If } \Tilde{s}_i[i'] \cap \Tilde{s}_j[j'] \neq \emptyset, \\[4pt]
0 & \text{Otherwise}.
\end{cases}
\label{eq:compatibility}
\end{equation}

For this purpose, we first encode an indeterminate string $\Tilde{s}$  as a Boolean matrix $\Tilde{s}^M$ that denotes the existence of a letter $c\in \Sigma$ at index $i'$. Hence, for all $1 \leq i \leq |\Tilde{s}|$ and $c \in \Sigma$, we define $\Tilde{s}^M[i',c]$ as:

\begin{equation}
\Tilde{s}^M[i',c] =
\begin{cases}
1 & \text{If } c \in \Tilde{s}[i'], \\[4pt]
0 & \text{Otherwise}.
\end{cases}
\label{eq:Xe-membership}
\end{equation}

After encoding strings $s_i$ and $s_j$ as Boolean matrices $\Tilde{s}_i^M$ and $\Tilde{s}_j^M$, the matrix $\mathcal{I}$ defined in Equation~\ref{eq:Xe-membership} is obtained by multiplying matrices $\Tilde{s}_i^M$ and $\Tilde{s}_j^M$, and then converting the result to binary by replacing all values greater than $1$ with $1$ (See Example~\ref{ex:indet} of the Appendix).

Without loss of generality, we assume that $|\Tilde{s}_i|$ and $|\Tilde{s}_j|$ is divisible by $\sigma$. We divide both the Boolean matrices $\Tilde{s}_i^M$ and $\Tilde{s}_j^M$ into square partitions, each having a size of $\sigma \times \sigma$ (a total of $\ell/\sigma$ partitions for each string). Thus, the encoding of the $\Tilde{s}_i$ and $\Tilde{s}_j$ requires $\bigo(|\Tilde{s}_i|\sigma)=\bigo(\sigma \ell)$ and $\bigo(|\Tilde{s}_j|\sigma)=\bigo(\sigma \ell)$ time and space, respectively. Square matrix multiplication of two matrices of size $\sigma \times \sigma$ can be achieved in $\bigo(\sigma^{w})$ time complexity, while the naive approach takes $\bigo(\sigma^3)$ time. We repeat this computation for all $(\ell/\sigma)^2$ blocks of size $\sigma \times \sigma$, and so $\bigo((\ell/\sigma)^2 \sigma^{w})=\bigo(\ell^2 \sigma^{w-2})$-time is needed to compute all blocks; that is, the entire $\mathcal{I}$ matrix. After computing the $\mathcal{I}$ matrix, $\bigo(\ell^2)$ additional time is required to compute $LCP^{H,k}_{(\Tilde{s}_i,\Tilde{s}_j)}$. Therefore, the total time required is $\bigo(\ell\sigma + \ell^2 \sigma^{w-2} + \ell^2) = \bigo(\ell\sigma + \ell^2 \sigma^{w-2})$. The encoded string matrices $\Tilde{s}_i^M$ and $\Tilde{s}_j^M$ require $\bigo(\sigma\ell)$ space, and so the $\mathcal{I}$ matrix requires $\bigo(\ell^2)$ space. Since $|\Tilde{s}_i|$ and $|\Tilde{s}_j|$ are both divisible by $\sigma$, we have $\sigma \leq \ell$. Thus, we get:

\begin{lemma}\label{lem:indetLCP}
    $LCP^{H,k}_{(\Tilde{s}_i,\Tilde{s}_j)}$ can be computed in $\bigo(\sigma\ell+\sigma^{w-2}\ell^2)$ time and $\ell^2$ space.
\end{lemma}

The same naive $\bigo(\sigma \ell^2)$ time complexity extends to the ELCS problem on multiple indeterminate strings. This can be achieved by expanding all indeterminate strings into their determinate realizations, whose total length is $\bigo(\sigma \ell^2)$. For this expanded string, we can construct a Generalized Suffix Tree (see~\cite[Section 7.7]{gusfield1997algorithms}). After a $\bigo(\sigma \ell^2)$ preprocessing step, constant-time $lca$ (longest common ancestor) queries are supported (see~\cite[Section 9.7]{gusfield1997algorithms}). This yields an overall time complexity of $\bigo(\sigma \ell^2)$ for computing ELCS. However, the efficient computation of ALCS also applies to ELCS by evaluating $LCP^{H,k}_{(\Tilde{s}_i,\Tilde{s}_j)}$ for $k=0$ (as noted in Section~\ref{sec:litReview}, the ELCS problem on multiple strings coincides with $Rkt$-LCS when $k=0$). Using Lemma~\ref{lem:indetLCP}, we obtain the following result:

\begin{theorem}
    The ELCS, Rk-LCS, Rkt-LCS of $\s{\Tilde{S}}=\{\Tilde{s}_1,\Tilde{s}_2,\cdots,\Tilde{s}_m\}$ for $\delta=H$ can be computed in $\bigo(m^2 \sigma\ell+\sigma^{w-2} N^2)$ time and $\bigo(m \ell^2)$ additional space.
\end{theorem}

Note that when $\sigma \leq 64$, we can encode indeterminate strings using the power-of-two encoding proposed in~\cite{btt2013} and use constant-time bit-wise operations to compute set intersections in the RAM model. This reduces the time required to compute the $LCP^{H,k}_{(\Tilde{s}_i,\Tilde{s}_j)}$ table to $\bigo(\ell^2)$ time.

\section{SETH-Hardness of \texorpdfstring{$Rk$-LCS}{Rk-LCS} \& \texorpdfstring{$Rk$-LCSS}{Rk-LCSS}}\label{sec:seth1}

In this section, we show the conditional lower bounds for the $Rk$-LCS and $Rk$-LCSS problems (defined in Section~\ref{sec:DefforApproxLCS}) for $\delta=H$ and for any $m>2$.

\begin{problem}
    [Orthogonal Vector (OV)] 
    Given a set $A \subseteq \{0, 1\}^d$, where $|A|=N_v$, determine whether there exists a pair of orthogonal vectors $u,v \in A$, such that $\Sigma^d_{p=1}u[p]v[p]=0$.
\end{problem}

\begin{problem}[$\mathcal{M}$-Orthogonal Vector ($\mathcal{M}$-OV)] Given $\mathcal{M}$ non-empty sets of vectors $X_1, X_2, \dots, X_{\mathcal{M}} \subseteq \{0, 1\}^d$, and $N_{\mathcal{M}} = \sum_{i=1}^{\mathcal{M}} |X_i|$, determine whether there exists a vector $u \in X_i$, for some $i \in [1..\mathcal{M}]$, and vectors $v_j \in X_j$, for all $j \in [1,\mathcal{M}] \setminus \{i\}$, such that $\Sigma^d_{p=1}u[p]v_j[p]=0$.
\end{problem}

In~\cite{WILLIAMS2005357} it has been shown that the OV problem does not have a strongly sub-quadratic time solution, unless SETH\footnote{The Strong Exponential Time Hypothesis (SETH) asserts that for every $\varepsilon > 0$, there exists an integer $q$ such that SAT on \textit{q}-CNF formulas with $m$ clauses and $n$ variables cannot be solved in $m^{O(1)}2^{(1-\varepsilon)n}$ time~\cite{WILLIAMS2005357}.} is false. Because $\mathcal{M}$-$OV$ generalizes $OV$, a subquadratic algorithm for $\mathcal{M}$-$OV$ (even for $\mathcal{M} = 2$) would imply a subquadratic algorithm for $OV$, contradicting the SETH-based lower bound. Therefore, unless SETH fails, $\mathcal{M}$-$OV$ cannot be solved in time $\bigo(N_v^{2 - \varepsilon})$ for any $\varepsilon > 0$, where $N_v = \sum_i |X_i|$. Hence, we have the following result:

\begin{theorem}\label{theo:movhard}
Suppose there is $\varepsilon > 0$ such that for all constants $c$, $\mathcal{M}$-$OV$ problem on a set of $N_v$ Boolean vectors of dimension $d = c\log N_v$ can be solved in $2^{o(d)}N^{2-\varepsilon}_v$ time. Then SETH is false.
\end{theorem}

Kociumaka et al.~\cite{Kociumaka2019LongestMismatches} demonstrate a $\bigo(\ell^2)$-time conditional lower bound for the \textit{Rk}-LCS problem for two strings ($m=2$) and $\delta=H$ by constructing two morphisms as shown in Figure~\ref{fig:morphismOne} in the appendix. We generalize this result to the \textit{Rk}-LCS problem for more than two strings ($m>2$) as follows: 
we apply the $\mu$ morphism to all the vectors in the set $X_1 \cup X_2 \cup \ldots \cup X_{\mathcal{M}}$ to construct string $s_1$. Then, we apply the $\tau$ morphism to sets $X_1, X_2, \dots, X_{\mathcal{M}}$ to obtain strings $s_2, s_3,\ldots,s_m$, respectively. This construction enables us to show that the \textit{Rk}-LCS problem has an $\bigo(m^2\ell^2)$ or $\bigo(N^2)$-time conditional lower bound using the same proof technique presented in Lemma 2.5 of~\cite{Kociumaka2019LongestMismatches}. Thus, we have the following result:

\begin{lemma}\label{lem:instance}
    Consider Boolean vectors $X_i=\{U^i_1,\ldots,U^i_{N_v/\mathcal{M}}\}$, $1 \leq i \leq \mathcal{M}$, each of dimension $d$, and $G = \gamma^d$, $\gamma= 100~1000$. The following string constructions are assigned to the strings of set $\s{S}=\{s_1,s_2,\ldots,s_m\}$:
    
\[
\begin{aligned}
& s_1=G^q\mu(U^1_1)\,G^q\mu(U^1_2)\,\ldots\,\mu(U^2_1)\,G^q\mu(U^2_2)\,\ldots\,\mu(U^\mathcal{M}_{N_v/\mathcal{M}})\,G^q\\
& s_2=G^q\tau(U^1_1)\,G^q\tau(U^1_2)\,\ldots\,\tau(U^1_{N_v/\mathcal{M}})\,G^q\\
\;\; &\vdots\\
& s_m=G^q\tau(U^\mathcal{M}_1)\,G^q\tau(U^\mathcal{M}_2)\,\ldots\,\mu(U^\mathcal{M}_{N_v/\mathcal{M}})\,G^q
\end{aligned}
\]
For $\mathcal{M}+1=m$, $q \geq 1$, and $k=d$:\\
(a) If $A$ contains $\mathcal{M}$-$OV$, then Rk-LCS for $\s{S}$ has length $\ell' \geq (14q+7)d$.\\
(b) If $A$ does not contain a $\mathcal{M}$-$OV$, then Rk-LCS for $\s{S}$ has length $\ell'' < (7q+14)d$.
\end{lemma}

\begin{theorem}\label{theo:rklcshard}
    Suppose there is a $\varepsilon > 0$ such that Rk-LCS for any $m>2$ can be solved in $\bigo(N^{2-\varepsilon})$ time on binary strings for $k=\Omega(\log N)$, then SETH fails.
\end{theorem}
\begin{proof}
    By Lemma~\ref{lem:instance}, we construct an instance of the $\mathcal{M}$-$OV$ problem with $N_v$ vectors of dimension $d$. This instance is equivalent to an instance of the $Rk$-LCS problem, where $k=d$ and $N = 7(m-1)d(2N_v/\mathcal{M}+1)+7d(2N_v+1)=\bigo(N_vd)$. Assuming that the $Rk$-LCS problem for any $m$ can be solved in $\bigo(N^{2-\varepsilon})$ time for $k=\Omega(\log N)$, we can solve the constructed instance of the $\mathcal{M}$-$OV$ problem in $\bigo(N^{2-\varepsilon}_vd^{O(1)})$ time for $d=c \log N_v$. This contradicts SETH. \QEDB
\end{proof}

A solution to the $Rk$-LCSS problem for $\delta=H$ is to construct a graph of $N$ vertices (representing all suffixes of $\s{S}$) and search for a clique of size $m$, where the suffixes originate from $m$ different strings and the longest common prefix between any two suffixes has at most $k$ mismatches. Clearly, this approach requires examining all $\binom{N}{m} = \bigo(N^m)$ possibilities.
We show that this time complexity is a conditional lower bound for this problem for any fixed $m$. For this, we introduce a new problem termed as the complete $\mathcal{K}$-$OV$ problem and show that it cannot be solved in $N^{\mathcal{K}-\varepsilon}_vpoly(d)$, unless SETH is false. Then we reduce it to the $Rk$-LCSS problem to establish the $\bigo(N^m)$ lower bound.

\begin{problem}[$\mathcal{K}$-$OV$~\cite{abboud2018more,manea2024subsequences}] Given $\mathcal{K}$ sets $X_1, \ldots, X_\mathcal{K} \subseteq \{0,1\}^d$, each consisting of $N_{v}$ Boolean vectors, find $x_1 \in X_1, \ldots, x_\mathcal{K} \in X_\mathcal{K}$, such that $\Sigma_{j=1}^d \Pi_{i=1}^\mathcal{K} x_i[j]=0$.
\end{problem}

\begin{lemma}
    \label{lem:kov}
    (\cite{manea2024subsequences}) $\mathcal{K}$-$OV$ problem cannot be solved in $N^{\mathcal{K}-\varepsilon}_vpoly(d)$ time for any $\varepsilon > 0$ and $d=\omega(\log N_v)$.
\end{lemma}

\begin{problem}
    [\textit{Complete} $\mathcal{K}$-$OV$] Given $\mathcal{K}$ sets $X_1, \ldots, X_\mathcal{K}$, each containing $N_v$ Boolean vectors, find $x_1 \in X_1, \ldots, x_\mathcal{K} \in X_\mathcal{K}$ such that $x_i$ and $x_j$ are orthogonal for any $1 \leq i, j \leq \mathcal{K}$, $i \neq j$.
\end{problem}

The \textit{Complete} $\mathcal{K}$-$OV$ problem is as hard as the $k$-Clique problem. An instance of \textit{Complete} $\mathcal{K}$-$OV$ is also an instance of the $\mathcal{K}$-$OV$ problem. Specifically, if a set $x_1 \in X_1, \ldots, x_\mathcal{K} \in X_\mathcal{K}$ is pairwise orthogonal, then $\sum_{j=1}^d \prod_{i=1}^\mathcal{K} x_i[j] = 0$, because for any index $j$, at most one $x_i[j]$ can be $1$; otherwise, at least two vectors are not orthogonal. Hence, if \textit{Complete} $\mathcal{K}$-$OV$ can be solved in $N^{k-\varepsilon}_vpoly(d)$, then $\mathcal{K}$-$OV$ can. Hence, we get the following result:

\begin{lemma}
    Complete $\mathcal{K}$-$OV$ cannot be solved in $N_v^{\mathcal{K}-\varepsilon}poly(d)$ time for any $\varepsilon > 0$ and $d=\omega(\log N_v)$.
\end{lemma}

As shown in Figure~\ref{fig:morphism2}, for the \textit{Rk}-LCSS problem, we construct $m$ distinct morphisms, each of length $2m+7$, denoted as $\tau_i$ for $1 \leq i \leq m$. The string gadget is defined as $\gamma = 1{0}^{m+2} \, 1{0}^{m+3}$, being distinct from all defined morphisms. Note that, for any $1 \leq i,j \leq m$ and $i \neq j$, $d_H(\tau_i(0), \tau_j(1)) = d_H(\tau_i(1), \tau_j(0))= d_H(\tau_i(0), \tau_j(0)) = 3$, $d_H(\tau_i(1), \tau_j(1)) = 5$. Moreover, the string $10^{m+3}$ has exactly two occurrences in $10^{m+3}a_1\ldots a_{m+3}10^{m+3}$ for $a_i \in \{0,1\}$ and $i \in \{1,...,m+3\}$. Therefore, analogous to Lemma~\ref{lem:instance}, we have the following results.

\begin{figure}
\centering
\begin{minipage}{0.48\textwidth}
\begin{align*}
\tau_i(0) &= 011 \; 0^{i-1}10^{m-i}\ \; \underline{10^{m+3}} \\
\tau_i(1) &= 000 \; 0^{i-1}10^{m-i}\ \; \underline{10^{m+3}} \\
\tau_j(0) &= 001 \; 0^{j-1}10^{m-j}\ \; \underline{10^{m+3}} \\
\tau_j(1) &= 111 \; 0^{j-1}10^{m-j}\ \; \underline{10^{m+3}} \\
\gamma &= 1{0}^{m+2} \; \underline{10^{m+3}}
\end{align*}
\end{minipage}
\hfill
\begin{minipage}{0.48\textwidth}
\begin{tikzpicture}
\node[draw, circle] (t_0) at (2, 0) {$\tau_i(0)$};
\node[draw, circle] (t'_0) at (5, 0) {$\tau_j(0)$};
\node[draw, circle] (t_1) at (2, -2) {$\tau_i(1)$};
\node[draw, circle] (t'_1) at (5, -2) {$\tau_j(1)$};
\draw (t_0) -- (t'_0) node[midway, above] {$3$};
\draw (t_1) -- (t'_1) node[midway, below] {$5$};
\draw (t'_0) -- (t'_1) node[midway, right] {$2$};
\draw (t_0) -- (t_1) node[midway, left] {$2$};
\draw (t_1) -- (t'_0) node[midway, right, xshift=6pt] {$3$};
\draw (t_0) -- (t'_1) node[midway, left, xshift=-6pt] {$3$};
\end{tikzpicture}
\end{minipage}
\caption{Hamming distances between different morphisms for $1 \leq i,j \leq m$, $i \neq j$ for $Rk$-LCSS problem.}
\label{fig:morphism2}
\end{figure}

\begin{lemma}\label{lem:instance2}
Consider sets $X_i=\{U_1^i,\ldots,U_{N_v}^i\}$, $1 \leq i \leq \mathcal{K}$ of Boolean vectors and $G = \gamma^d$. The strings are constructed as follows, where $1\leq i\leq m$, $\s{S}=\{s_1,s_2,\ldots,s_m\}$, and $\mathcal{K}=m$:
$$s_{i} =G^{q}\tau_{i}\bigl(U_{1}^{i}\bigr)G^{q}\tau_{i}\bigl(U_{2}^{i}\bigr)\ldots\tau_{i}\bigl(U_{N_v}^{i}\bigr)G^{q}$$ 
For some $q \geq 1$ and $k=3d$:\\
(a) If sets $X_1,\ldots,X_\mathcal{K}$ contain a \textit{Complete} $\mathcal{K}$-$OV$, then Rk-LCSS under Hamming distance for $\s{S}$ is a set in which each substring has a length of $\ell' \geq (2m+7)(2q+1)d$.\\
(b) If sets $X_1,\ldots,X_\mathcal{K}$ do not contain a \textit{Complete} $\mathcal{K}$-$OV$, then Rk-LCSS under Hamming distance for $\s{S}$ is a set in which each substring has a length of $\ell'' < (2m+7)(q+2)d$.\\
\end{lemma}

\begin{theorem}
    Suppose there exists $\varepsilon > 0$ such that Rk-LCSS can be solved in $\bigo((N_v/m^2)^{m-\varepsilon})$ time for any $m \geq 2$ on binary strings for $k=\omega(\log N)$. Then SETH is false.
\end{theorem}
\begin{proof}
By Lemma~\ref{lem:instance2} with $q=1$ and $\mathcal{K}=m$, we construct an instance of \textit{Complete} $\mathcal{K}$-$OV$ problem with the total of $N_v$ vectors of dimension $d$ distributed equally across $m$ sets. It is an equivalent instance of $Rk$-LCSS problem, where $k=3d$ and the total length of strings is $N=m(2N_v+1)(2m+7)d=\bigo(N_vm^2d)$. Therefore, if the $Rk$-LCSS problem can be solved in $\bigo((\frac{N}{m^2})^{m-\varepsilon})$ time for $k=\omega(\log N)$, then the constructed instance of the \textit{Complete} $\mathcal{K}$-$OV$ problem can be solved in $\bigo(N_v^{m-\varepsilon}d^{O(1)})$ time for $d=c \log N_v$. This contradicts SETH. \QEDB
\end{proof}

\begin{corollary}
    Suppose there exists $\varepsilon > 0$ such that Rk-LCSS can be solved in $\bigo(N^{m-\varepsilon})$ time for a fixed number of strings in $\s{S}$ on binary strings for $k=\omega(\log N)$. Then SETH is false.
\end{corollary}

\section{Future works \& Open Problems}
As future work, it would be worthwhile to implement the polylogarithmic-time approaches and empirically compare their performance with quadratic-time solutions. Another promising direction is the development of algorithms for these problems on elastic-degenerate strings. Establishing a conditional lower bound of $\bigo(N^m)$ for the \textit{Rk}-LCSS problem for any $m$ remains an open problem. It would also be interesting to investigate strongly subquadratic-time solutions in a quantum setting. Finally, it is of interest to explore improved time complexities for the ELCS problem on multiple indeterminate strings.

\begin{credits}
\subsubsection{\ackname} 
The second and third authors are funded by the Natural Sciences \& Engineering Research Council of Canada [Grant Numbers RGPIN-2024-06915 and RGPIN-2024-05921], respectively. 
The authors wish to thank Dr. Brian Golding for suggesting this research topic and guiding us during the problem formulation. 
\end{credits}

\bibliographystyle{splncs04}
\bibliography{references}

\appendix
\section{Appendix: Examples}
\begin{table}
\centering
\caption{$LCP^{H,2}_{(s_1,s_2)}$ table for $s_1=GTACAAT$ and $s_2=CTTGTA$. For example,
$LCP^{H,2}_{(s_1,s_2)}[2,3]=4$ is the length of the longest commons prefix of suffixes $s_1[2..|s_1|]=TACAAT$ and $s_2[3..|s_2|]=TGTA$ up to $2$ mismatches.} \label{fig:lcpkexample}
\renewcommand{\arraystretch}{1.1} 
\setlength{\tabcolsep}{4pt} 
\begin{tabular}{|c|c|c|c|c|c|c|}
\hline
 &  C &  T &  T &  G &  T &  A \\ \hline
 G &  3 &  3 &  2 &  3 &  2 &  1 \\ \hline
 T &  2 &  3 &  4 &  2 &  2 &  1 \\ \hline
 A &  2 &  2 &  2 &  3 &  2 &  1 \\ \hline
 C &  3 &  2 &  2 &  3 &  2 &  1 \\ \hline
 A &  3 &  2 &  3 &  2 &  2 &  1 \\ \hline
 A &  2 &  2 &  2 &  2 &  2 &  1 \\ \hline
 T &  1 &  1 &  1 &  1 &  1 &  1 \\
\hline
\end{tabular}
\end{table}

\begin{table}
\centering
\caption{The $lengthStat^{H,1}_{(3,1)}$ table for $\s{S}=\{TTGAC,CGAAAT,TGGTA\}$, where $k=1$. The $lengthStat^{H,1}_{(3,1)}[3,2]=1$ indicates the 1-approximate occurrence of the prefix of length three of $s_1[3..5]$ $(GAC)$, somewhere in $s_2$ $(s_2[2..4]=GAA)$.}
\label{tab:lengthstat}
\renewcommand{\arraystretch}{1.1} 
\setlength{\tabcolsep}{4pt}
\begin{tabular}{|c|c|c|c|c|}
\hline
 & 1 ($s_1$) & 2 ($s_2$) & 3 ($s_3$) & 4 (Frequency) \\ \hline
1 & 1 & 1 & 1 & 3 \\ \hline
2 & 1 & 1 & 1 & 3 \\ \hline
3 & 1 & 1 & 0 & 2 \\ \hline
\end{tabular}
\end{table}

\begin{figure}\label{fig:strmorph}
\centering
\begin{minipage}{0.48\textwidth}
\begin{align*}
\mu(0) &= 011 \; \underline{1000} \\
\mu(1) &= 000 \; \underline{1000} \\
\tau(0) &= 001 \; \underline{1000} \\
\tau(1) &= 111 \; \underline{1000} \\
\gamma &= 100 \; \underline{1000}
\end{align*}
\end{minipage}
\hfill
\begin{minipage}{0.48\textwidth}
\begin{tikzpicture}
\node[draw, circle] (mu_0) at (2, 0) {$\mu(0)$};
\node[draw, circle] (t_0) at (5, 0) {$\tau(0)$};
\node[draw, circle] (mu_1) at (2, -2) {$\mu(1)$};
\node[draw, circle] (t_1) at (5, -2) {$\tau(1)$};
\draw (mu_0) -- (t_0) node[midway, above] {$1$};
\draw (mu_1) -- (t_1) node[midway, below] {$3$};
\draw (t_0) -- (t_1) node[midway, right] {$2$};
\draw (mu_0) -- (mu_1) node[midway, left] {$2$};
\draw (mu_1) -- (t_0) node[midway, right, xshift=6pt] {$1$};
\draw (mu_0) -- (t_1) node[midway, left, xshift=-6pt] {$1$};
\end{tikzpicture}
\end{minipage}
\caption{Hamming distances between morphisms~\cite{Kociumaka2019LongestMismatches}. All underlined segments are the fixed part and common to every morphism.}
\label{fig:morphismOne}
\end{figure}

\begin{example}\label{ex:indet}
Assume $\Tilde{s}_i=[A,T]G[CG]T$, $\Tilde{s}_j=C[A,T]TA$, and $\Sigma=\{A,C,G,T\}$.
\[
\Tilde{s}_i^M =
\begin{bmatrix}
1 & 0 & 0 & 1 \\
0 & 0 & 1 & 0 \\
0 & 1 & 1 & 0 \\
0 & 0 & 0 & 1
\end{bmatrix}
\quad
\Tilde{s}_j^M =
\begin{bmatrix}
0 & 1 & 0 & 0 \\
1 & 0 & 0 & 1 \\
0 & 0 & 0 & 1 \\
1 & 0 & 0 & 0
\end{bmatrix}
\quad
\Tilde{s}_i^M \cdot \Tilde{s}^{M}_j=
\begin{bmatrix}
0 & 2 & 1 & 1 \\
0 & 0 & 0 & 0 \\
1 & 0 & 0 & 0 \\
0 & 1 & 1 & 0
\end{bmatrix}
\quad
\mathcal{I}=
\begin{bmatrix}
0 & 1 & 1 & 1 \\
0 & 0 & 0 & 0 \\
1 & 0 & 0 & 0 \\
0 & 1 & 1 & 0
\end{bmatrix}
\]
\end{example}

\end{document}